\documentclass[review]{elsarticle}

\usepackage{lineno,hyperref}
\modulolinenumbers[5]
\usepackage{graphicx}
\usepackage{subcaption}
\usepackage{amssymb}
\usepackage{amsmath}
\usepackage{verbatim}

\newcommand{\RNum}[1]{\uppercase\expandafter{\romannumeral #1\relax}}

\begin{document}

\begin{frontmatter}

\title{A method to compute the communicability of nodes through causal paths in temporal networks}

\author{Agostino Funel}
\address{ENEA, Via E. Fermi, 1 - 80055 Portici (Naples), Italy}




\begin{abstract}
We present a method aimed to compute the communicability (broadcast and receive) of nodes through causal paths in temporal networks. 
The method considers all possible combinations of chronologically ordered products of adjacency matrices of the network snapshots and by means of a damping procedure favors the paths that have high communication efficiency.
 We apply the method to four real-world networks of face-to-face human contacts and  identify the nodes with high communicability. The accuracy of the method is proved by studying the spread of an epidemic in the networks using the susceptible-infected-recovered model. We show that if a node with high broadcast is chosen as the origin of the outbreak of infection then the epidemic spreads early while it is delayed and inhibited if the origin of infection is a node with low broadcast. Receiving nodes can be treated as broadcasters if the arrow of time is reversed.


\end{abstract}

\begin{keyword}
\texttt{temporal networks, causal paths, communicability, epidemic spreading} 
\end{keyword}

\end{frontmatter}


\section{Introduction}\label{sec_intro}

\begin{figure}[!htbp]
\begin{center}
    \includegraphics[width=\textwidth]{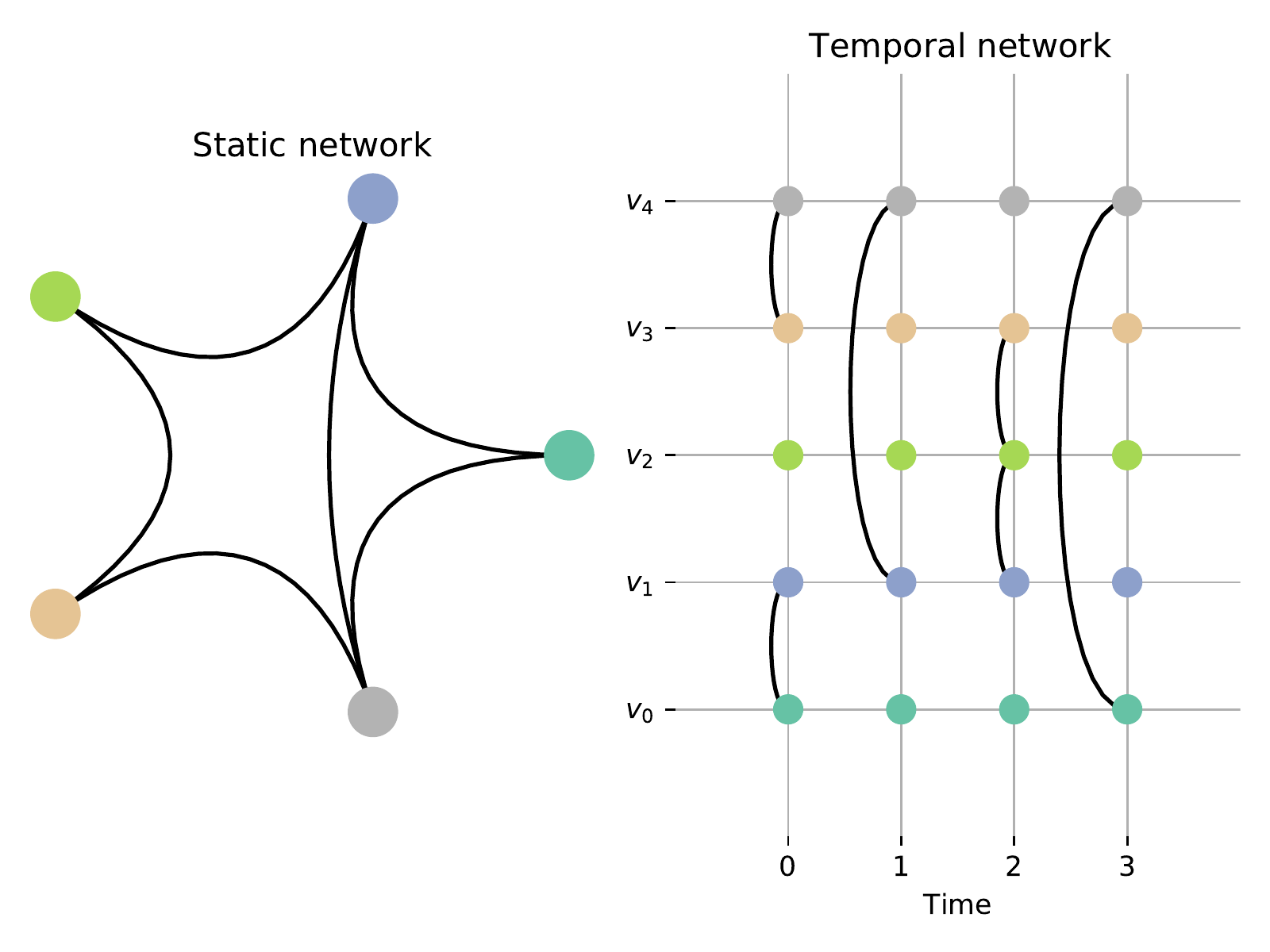}
\caption{A toy network composed of five nodes. The static aggregate network is obtained by connecting two nodes if they interacted at least once during the period of observation. The temporal network shows the active links for each of the four network snapshots.}
\label{fig_toy_netw}
\end{center}
\end{figure}

The problem of identifying nodes with high communicability in a temporal network, that is, nodes with a great ability to broadcast and receive something that can spread in the network, is interesting not only from a theoretical perspective but also for practical aspects. 
In a temporal network the nodes interact with each other over time and the type of interaction depends on the nature of the network.
 When many nodes interact during a certain period of time all links which traverse them form pathways and, even if at each interaction any two nodes may communicate, the arrow of time induces a direction in the communication flow from a source to a target. For example, looking at Fig.~\ref{fig_toy_netw} it is clear that in the time interval $[0,2]$ communication can flow from $v_0$ to $v_2$ but not in the opposite direction and this cannot be inferred from the static aggregate network. 

A node with high ability to broadcast and receive should be able to communicate efficiently with as many other nodes as possible and this occurs if the node is a hub crossed by most of the communication paths, or can be connected to other nodes through short paths. These ideas led to the concept of node centrality~\cite{Katz_1953,Freeman_1979,Freeman_1977} widely applied to social analysis. Many algorithms have been developed to compute node centrality in the case of static networks, while in the case of temporal networks the situation is more complex  because the set of paths between nodes varies during the evolution of the network and causality needs to be considered. We use the notation $(v_i, v_j, t)$ to indicate the existence of a link between $v_i$ and $v_j$ at time $t$, regardless of direction. Causality implies that the chronological order must be respected, that is, if exist links $(v_i, v_j, t_j)$ and $(v_j, v_k, t_k)$ then a causal path from $v_i$ to $v_k$ exists only if $t_j \leq t_k$. The length $\ell$ of a causal path $(v_0, v_1, t_1)(v_1, v_2, t_2)\cdot \cdot \cdot (v_{l-1}, v_{\ell}, t_{\ell})$, where $t_1 \leq \cdot \cdot \cdot \leq t_{\ell}$, is the number of links that traverse its nodes. The duration of the path is $\delta = t_{\ell} - t_1$ and a single link can be considered as an instantaneous event. A node could be traversed several times, giving rise to topological loops. Here the communication flows from node $v_0$ to node $v_{\ell}$. Note that if the arrow of time is reversed the communication will flow in the opposite direction and the roles of broadcast and receive nodes will be swapped. 
Considering again the toy network of Fig.~\ref{fig_toy_netw}, we see that node $v_2$ can broadcast only to nodes $v_1$ and $v_3$ and can receive from all other nodes, while node $v_1$ can communicate (broadcast and receive) with all other nodes. Hence, the communicability of node $v_1$ is higher than that of node $v_2$. 
In this paper we present a method aimed to compute the communicability in temporal networks. The method can be applied to many cases of practical interest. For example, in a social network nodes that have high communicability are individuals who may influence the diffusion of information and opinions; in an economic network they are shareholders who may affect the stock market. Nodes that have high broadcast can quickly trigger and speed up processes that may spread in the network. The identification of nodes with high communicability does not depend on the particular diffusive process, but only on the links that are active during the evolution of the network.


In this work we compute the communicability of four real-world temporal networks of human proximity contacts. The accuracy of the method is proved by studying the spread of an epidemic in the networks using the susceptible-infected-recovered (SIR) model. We show that if a node with high broadcast is chosen as the origin of the infection then the epidemic spreads early while it is delayed and inhibited in the case a node with low broadcast is chosen as the seed of infection. The same behavior of the epidemic trend is observed if the diffusive process occurs on the temporal network obtained by reversing the arrow of time and the role of broadcast is replaced by that of receive for the node that is the origin of the infection. 

The paper is organized as follows: in Sec.~\ref{sec_relwork} we present a previous work in which the computation of the communicability matrix is based on the extension to temporal networks of a centrality measure used for static networks; in Sec.~\ref{sec_comm} we present a method to compute the communicability which privileges causal paths that have high communication efficiency; in Sec.~\ref{sec_app} we compute the communicability of four real-world temporal networks of human proximity contacts and prove the accuracy of the used method by studying the spread of an epidemic in the networks. Finally, in Sec.\ref{sec_conc} we report the main conclusions.

\section{Related work}\label{sec_relwork}
In Ref.~\cite{Grindrod_2011} authors computed the communicability by extending to temporal networks a measure of centrality proposed by Katz~\cite{Katz_1953} for static networks. A network $G = (V, E)$ composed of $N = |V|$ nodes $v_i \in V$ and $|E|$ edges can be described by its $N \times N$ adjacency matrix matrix $\mathbf{A}$ defined as $\mathbf{A}_{ij} = 1$ if nodes $(v_i,v_j)$ are connected, $\mathbf{A}_{ij} = 0$ otherwise. The element of the $n$-power $\mathbf{A}^{n}_{ij}$ is the number of paths of length $n$ connecting nodes $(v_i,v_j)$. The sum $ \mathbf{C} = \mathbf{I} + \sum_{n = 1}^{\infty} a^n \mathbf{A}^n$, where $\mathbf{I}$ is the identity matrix, converges to the resolvent $(\mathbf{I} -a \mathbf{A})^{-1}$ if $a < \rho(\mathbf{A})^{-1}$, where $\rho(\mathbf{A})$ is the spectral radius of $\mathbf{A}$, that is, the maximum eigenvalue in modulus. Thus the element $\mathbf{C}_{ij}$ counts the number of paths of any length connecting nodes $v_i$ and $v_j$ where each path is weighted by the factor $a^{\ell}$ being $\ell$ the length of the path. A temporal network can be represented by a sequence of chronologically ordered snapshots $\{ \mathbf{A}^{[0]},\mathbf{A}^{[1]}, ...,\mathbf{A}^{[T]} \}$ collected during the period of observation $[t_0, t_T]$. Here $\mathbf{A}^{[k]}$ is the adjacency matrix of the network at the instant $t_k \in \{t_0,...,t_T\}$. The communicability matrix proposed in Ref.~\cite{Grindrod_2011} is $\mathbf{\mathcal{C}} = \prod_{n = 0}^{T} (\mathbf{I} -a \mathbf{A}^{[n]})^{-1}$. Similarly to the static case the element $\mathbf{\mathcal{C}}_{ij}$ is the number of causal paths of any length $\ell$ connecting nodes $v_i$ and $v_j$ and each causal path is weighted by the factor $a^{\ell}$. Because the arrow of time dictates the direction of the communication flow, the communicability matrix $\mathbf{\mathcal{C}}$ is not symmetric even if that were the matrices $\mathbf{A}^{[k]}$. Hence, $\mathbf{\mathcal{C}}_{ij}$ indicates of how well node $v_i$ may transmit information to node $v_j$. The broadcast and receive of a node $v_i$ are defined as the row and column sums $B(i) = \sum_{k = 1}^{N} \mathbf{\mathcal{C}}_{ik}$ and $R(i) = \sum_{k = 1}^{N} \mathbf{\mathcal{C}}_{ki}$. We would like to point out that the path weight does not depend on time, which is not beneficial if we want to privilege  paths that favor communication between the nodes as soon as possible after the initial instant $t_0$ and in the shortest time. As an example, consider the two sets of paths with the same length associated to the products $\mathbf{A}^{[0]}\mathbf{A}^{[1]}\mathbf{A}^{[2]}$ and $\mathbf{A}^{[T-2]}\mathbf{A}^{[T-1]}\mathbf{A}^{[T]}$ contributing to the communicability matrix. Although the same weight is associated to both sets, the effects on the communication caused by paths belonging to the first set will occur from the initial instant while paths belonging to the other set will contribute to the communication only towards the end of the period of observation.
 Another point to consider with this approach is that the correctness of the computation of the communicability matrix is guaranteed only if $a < \rho(\mathbf{A}^{[k]})^{-1} \, \forall k$, which implies the calculation of the spectral radius for all adjacency matrices, an effort that may be computationally infeasible if the number of adjacency matrices and their size grows too much.

Methods based on the matrix exponential have been proposed. In Ref.~\cite{Estrada_2013} the temporal network is described as a quantum system and the communicability matrix corresponds to the imaginary-time propagator of a quantum random walk in the network.

\section{Communicability}\label{sec_comm}
It is reasonable to suppose that a good communicability should allow the broadcast and reception of information between any pair of nodes as soon as possible after the initial instant $t_0$ of the period of observation $[t_0, t_T]$ of the network, in the shortest time, and using the fewest intermediate nodes. Since our considerations depend neither on the initial instant $t_0$ nor on the duration $T$ of the period of observation we assume, without loss of generality, that $t_0 = 0$ and $t_T = T$. Moreover, we also assume that the adjacency matrices describing the temporal network $\{ \mathbf{A}^{[0]},\mathbf{A}^{[1]}, ...,\mathbf{A}^{[T]} \}$ are collected in strict chronological order $0 = t_0 < t_1 < \cdot \cdot \cdot < t_T = T$. A causal path $(v_0, v_1, t_{i_1})(v_1, v_2, t_{i_2}) \cdot \cdot \cdot (v_{\ell-1}, v_{\ell}, t_{i_{\ell}})$ of length $\ell$ and duration $\delta = t_{i_{\ell}} - t_{i_1} \leq T$ is defined, in addition to its origin $v_0$ and its links, also by its initial instant $t_{i_1}$ and a chronologically ordered sequence of times $t_{i_1} < t_{i_2} < \cdot \cdot \cdot < t_{i_{\ell}}$, where $t_{i_n} \in \{t_0, t_1, \cdot \cdot \cdot, t_T \} \, \forall t_{i_n} \mid n = \{1,...,\ell\}$. Therefore, the communication efficiency of a path increases the more its initial instant $t_{i_1}$ is close to $t_0$ and the shorter its duration $\delta$ and length $\ell$. In order to privilege those paths  that have the above properties we propose to compute the communicability matrix as:

\begin{equation}
\mathbf{\mathcal{C}}^{(a)} = \prod_{n = 0}^{T} (\mathbf{I} + a^{w_n} \mathbf{A}^{[n]})
\label{eq_defcmtx}
\end{equation}

where $0 < a < 1$ and $a^{w_n}$ is a damping factor that depends on time through a weight function $w_n \triangleq w(t_n)$ which is non-negative $w_n \geq 0 \, \forall t_n$ and monotonically increasing:  $w_m < w_n \, \forall m,n \mid t_m < t_n$.  

The communicability matrix above yields all possible combinations of chronologically ordered products of adjacency matrices

\begin{equation}
\begin{split}
\mathbf{\mathcal{C}}^{(a)} & = \mathbf{I} + a^{w_0} \mathbf{A}^{[0]} + a^{w_1} \mathbf{A}^{[1]} + a^{w_2} \mathbf{A}^{[2]} + \cdot \cdot \cdot + a^{w_T} \mathbf{A}^{[T]} + \\
& + a^{w_0}a^{w_1} \mathbf{A}^{[0]}\mathbf{A}^{[1]} + a^{w_0}a^{w_2} \mathbf{A}^{[0]}\mathbf{A}^{[2]} + a^{w_1}a^{w_2} \mathbf{A}^{[1]}\mathbf{A}^{[2]} + \cdot \cdot \cdot \\
& + a^{w_0}a^{w_1}a^{w_2} \mathbf{A}^{[0]}\mathbf{A}^{[1]}\mathbf{A}^{[2]} + \cdot \cdot \cdot + a^{w_0}a^{w_1}a^{w_2}\cdot \cdot \cdot a^{w_T}\mathbf{A}^{[0]}\mathbf{A}^{[1]}\mathbf{A}^{[2]} \cdot \cdot \cdot \mathbf{A}^{[T]}
\end{split}
\label{eq_defcmtxexp}
\end{equation}

 and, by construction, the damping factors favor causal paths with high communication efficiency by giving them a higher weight, and suppress the less efficient ones. 
The weight function should not increase to fast to prevent an underestimations of paths that start not too long after the initial instant and may give a non negligible contribution to the communicability. On the other hand, a weight function that increases to slowly may give rise to an overestimation of inefficient paths. We heuristically found that a suitable choice is the logarithm function and adopt $w_n = \ln(t_n + 1)$ to make the weight function well defined at $t_0 = 0$. The element $\mathbf{\mathcal{C}}^{(a)}_{ij}$ is a weighted sum of all possible causal paths from node $v_i$ to node $v_j$ that can be observed during the period of observation of the network. For a given value of $a$ the broadcast $B(i,a)$ and receive $R(i,a)$ of a node $v_i$ can be computed as the row and column sums of $\mathbf{\mathcal{C}}^{(a)}$, respectively. Because the number of all possible causal paths becomes huge as the number of network snapshots increases, to avoid overflow in the computation we use the Euclidean normalized version $\hat{\mathbf{\mathcal{C}}}^{(a)} = \mathbf{\mathcal{C}}^{(a)} / \parallel \mathbf{\mathcal{C}} \parallel_2$ of the communicability matrix:

\begin{equation}
B(i,a) = \sum_{k = 1}^{N} \hat{\mathbf{\mathcal{C}}}^{(a)}_{ik}, \; R(i,a) = \sum_{k = 1}^{N} \hat{\mathbf{\mathcal{C}}}^{(a)}_{ki}.
\label{eq_bra}
\end{equation}

In the limit $a \rightarrow 0$ the damping procedure may suppress too much those paths that occur in a temporal interval that contains many snapshots and whose extremes are not too close to either $t_0$ or $t_T$. In the limit $a \rightarrow 1$ all paths have roughly the same weight and the identification of paths with high communication efficiency loses effectiveness. We thus divide the range of $a$ into $K = 1 / \epsilon$ sub-intervals of the same width $\epsilon$ (in this work we set $\epsilon = 0.1$), compute the broadcast and receive for each value $a_n = n \epsilon, \, n = 1,2,...,M = K -1$ and  use the average:

\begin{equation}
B(i) = \frac{1}{M}\sum_{n = 1}^{M} B(i, a_n), \; R(i) = \frac{1}{M}\sum_{n = 1}^{M} R(i, a_n).
\label{eq_br}
\end{equation}

The case $a = 1$ was considered in Ref.~\cite{Lentz_2013} to study the distribution of shortest path duration's and to quantify the goodness of the static time-aggregated representation of a temporal network. We note from Eq.~\ref{eq_bra} that the roles of the broadcast and receive are swapped if we use the transpose of the communicability matrix. From Eq.~\ref{eq_defcmtxexp} we observe that computing the transpose of $\mathbf{\mathcal{C}}^{(a)}$ is equivalent to reverse the chronological order of the network snapshots. This allows us to analyze the propensity of a node to receive information or to be a target in a diffusive process which evolves in the network considering it as a broadcaster when the diffusion occurs in the temporal network obtained by reversing the arrow of time $\{ \mathbf{A}^{[0]},\mathbf{A}^{[1]}, ...,\mathbf{A}^{[T]} \} \rightarrow \{ \mathbf{A}^{[T]},\mathbf{A}^{[T-1]}, ...,\mathbf{A}^{[0]} \}$.

The method that we have proposed to compute the communicability has the advantage that it can be easily implemented on a computing system because it is based on matrix multiplication and there are many very efficient software packages designed for this purpose. In this work we used Armadillo~\cite{Sanderson_2016, Sanderson_2018}, a high quality C++ linear algebra library.

\section{Applications}\label{sec_app}
In this section we analyze the communicability of four real-world temporal networks of human contacts. The datasets are publicly available and are provided by the SocioPatterns~\cite{sociopatterns} collaboration. Contacts data was collected by using wearable badges equipped with Radio Frequency Identification (RFID) devices. The acquisition system ensured a level of face-to-face proximity ($\sim$1.5 meter) with a temporal resolution of 20 seconds~\cite{rfid}. The networks analyzed in this work are undirected and refer to different social contexts: \textbf{Conference}: contacts between participants to a scientific conference~\cite{sfhhwp2}; \textbf{School}: contacts between students in a high school~\cite{highschool}; \textbf{Hospital}: contacts between patients and healthcare personnel in a hospital ward~\cite{hospital}; \textbf{Workplace}: contacts between employees in an office building~\cite{sfhhwp2}. 
Table~\ref{tab_datasets} shows the size and temporal information of the networks.

\begin{table*}[h!]
\begin{center}
\begin{tabular}{|c|c|c|c|}
\hline
Network &       Nodes   &       Snapshots      &       Observation duration [h]  \\
\hline
\hline
Conference            &       403     & 3509 &       19.5  \\
\hline
School     &       180     & 11273 &       62.6  \\
\hline
Hospital        &       75      & 9453 &       52.5  \\
\hline
Workplace       &       217     & 18488 &       102.7   \\
\hline
\end{tabular}
\end{center}
\caption{Temporal networks of human proximity contacts analyzed in this work.}
\label{tab_datasets}
\end{table*}

\begin{figure*}
  \begin{subfigure}[b]{0.5\textwidth}
    \includegraphics[width=\textwidth]{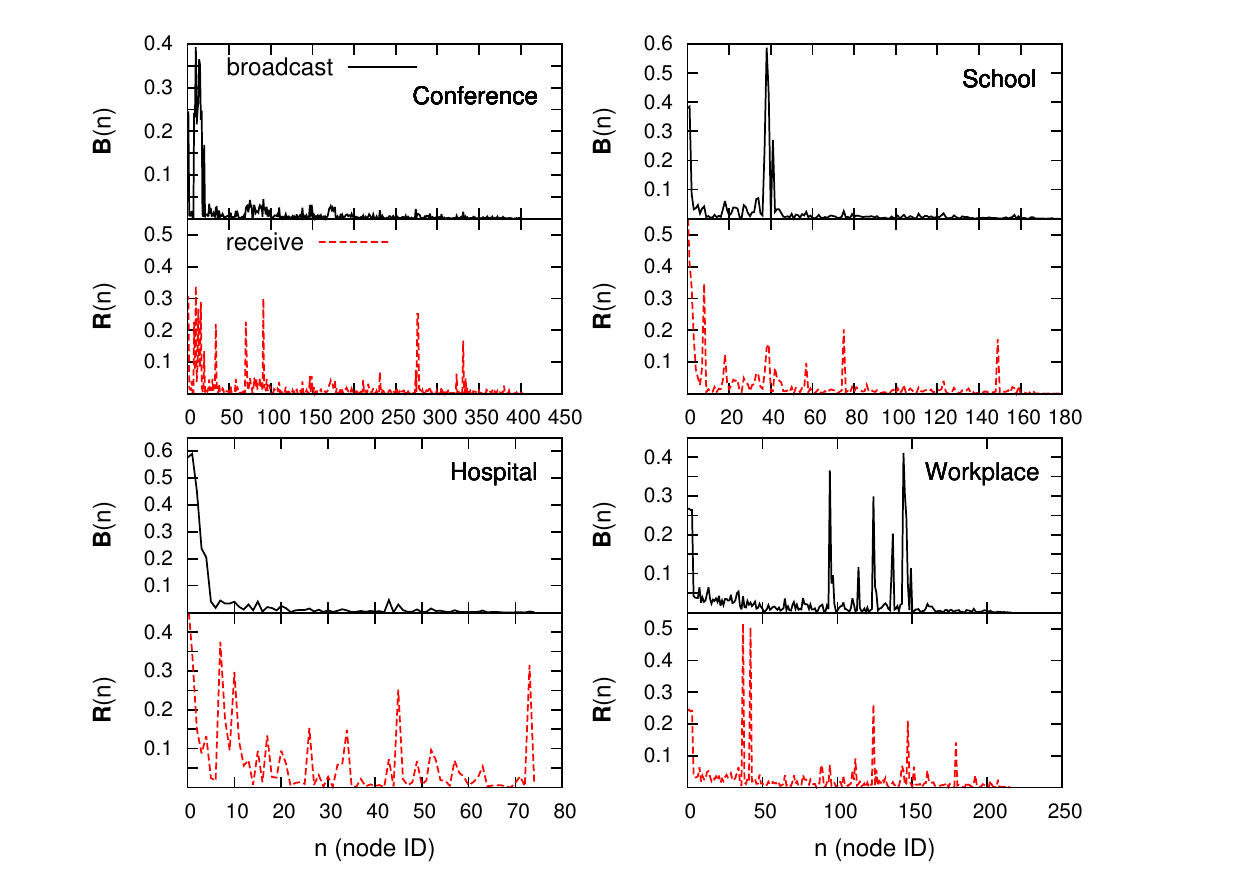}
    \label{fig_brvn}
  \end{subfigure}
  \begin{subfigure}[b]{0.5\textwidth}
    \includegraphics[width=\textwidth]{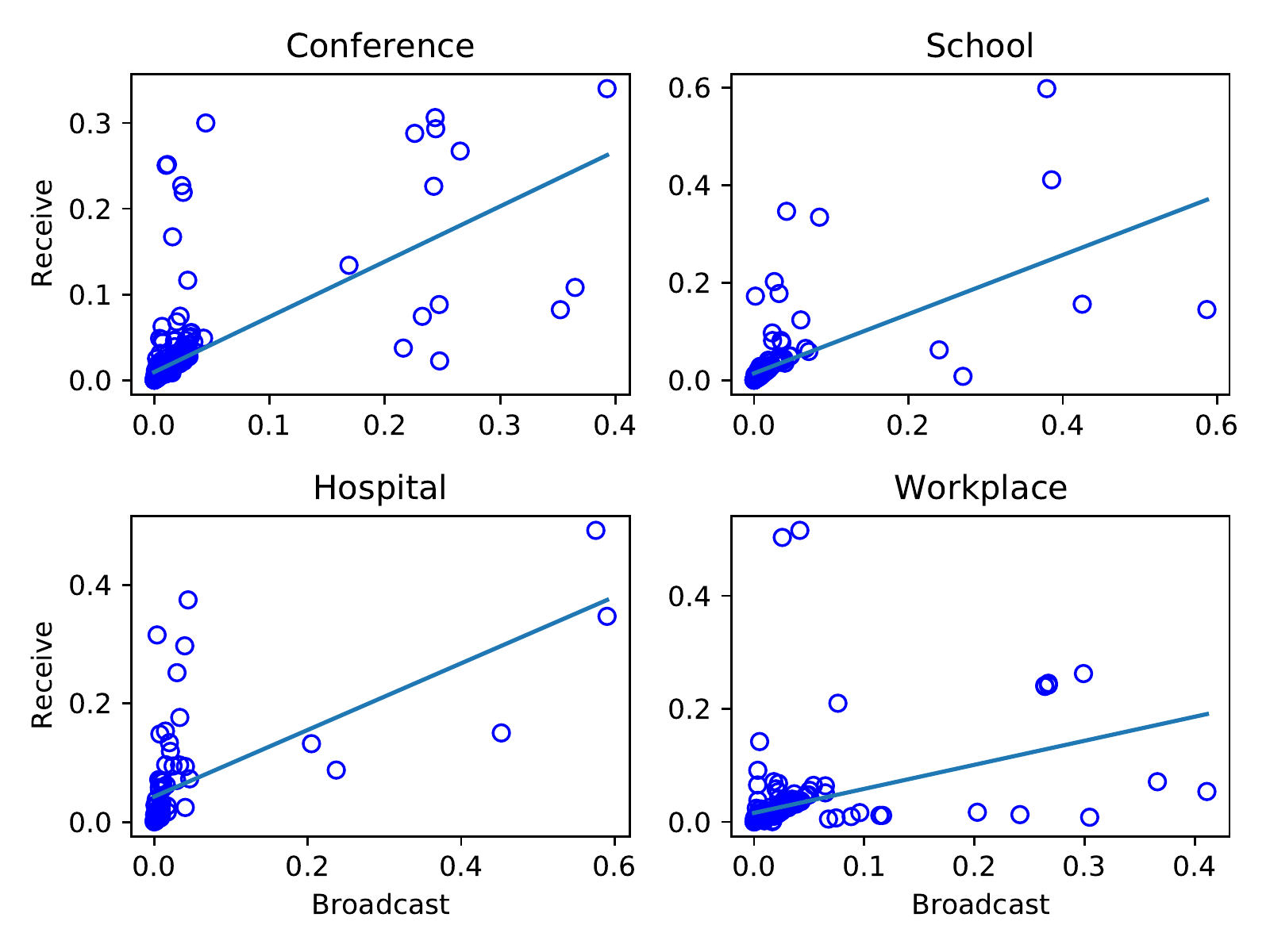}
    \label{fig_corrbr}
  \end{subfigure}
\caption{\textbf{Left}: broadcast (B) and receive (R) of the nodes of the analyzed networks. The peaks clearly identify the nodes that have high communicability. \textbf{Right}: broadcast and receive scatter plots. The figure also shows the best fit line. The Pearson correlation coefficients for the Conference, School, Hospital and Workplace networks are $r = \{0.67, 0.62, 0.64, 0.42 \}$, respectively.}
\label{fig_brcorrbr}
\end{figure*}

Fig.~\ref{fig_brcorrbr} shows the (normalized) broadcast and receive of the nodes of the analyzed networks. The nodes are labeled with integer numbers $n = 0,1,...,N-1$ being $N$ the size of the network. The peaks clearly identify the nodes that have high communicability. A node that has high broadcast but low receive can be considered mainly a source of information rather than a collector during the evolution of the network. The opposite can be said for a node  that has high receive. A node for which the ratio $B/R \simeq 1$ broadcasts and receives, on average, in a balanced way. The same figure also shows the broadcast and receive scatter plots. For all datasets there is a moderate positive correlation.

In order to prove the accuracy of the method we study the spread of epidemics in the networks.  We consider the discrete time susceptible-infected-recovered (SIR) model. In Ref.\cite{Koher_2019} are described two approaches to the SIR model for temporal networks. In the first one probabilities are defined only at the node level while in the other probabilities are defined also at the edge level. The reader is encouraged to read the work for more details. Here we focus on the first approach, which is more intuitive,  but we also performed simulations of the the edge-based model on the datasets and for both approaches we came to the same conclusions.  We denote, at any time step $t$, the probability for a node $v_k$ to be susceptible, infected or recovered as $S_k(t)$, $I_k(t)$, $R_k(t)$, respectively, and the adjacency matrix as $\mathbf{A}^{[t]} \triangleq \mathbf{A}(t)$. A susceptible node can contract the disease through contact with an infected neighbor with probability $\alpha$. An infected node can recover to a state of permanent immunity with probability $\beta$. The dynamic equations of the model are:

\begin{equation}
S_k(t+1) = S_k(t) \prod_{n \in V} [1 - \alpha \mathbf{A}_{nk}(t) I_n(t) ]
\label{eq_sir1}
\end{equation}

\begin{equation}
I_k(t+1) = (1 - \beta) I_k(t) + S_k(t) \{ 1 - \prod_{n \in V} [ 1 - \alpha \mathbf{A}_{nk}(t) I_n(t) ] \}
\label{eq_sir2}
\end{equation}

and the recover probability follows from the conservation equation $S_k(t) + I_k(t) + R_k(t) = 1$. The model is complete by assigning the initial conditions $S_k(0) = s_0$, $I_k(0) = 1 - s_0$ and $R_k(0) = 0$.


We expect that if we choose a node with high broadcast as the origin of infection the epidemic will spread earlier and on a larger scale than if we choose a node with low broadcast.

\begin{figure}[!htbp]
\begin{center}
    \includegraphics[width=\textwidth]{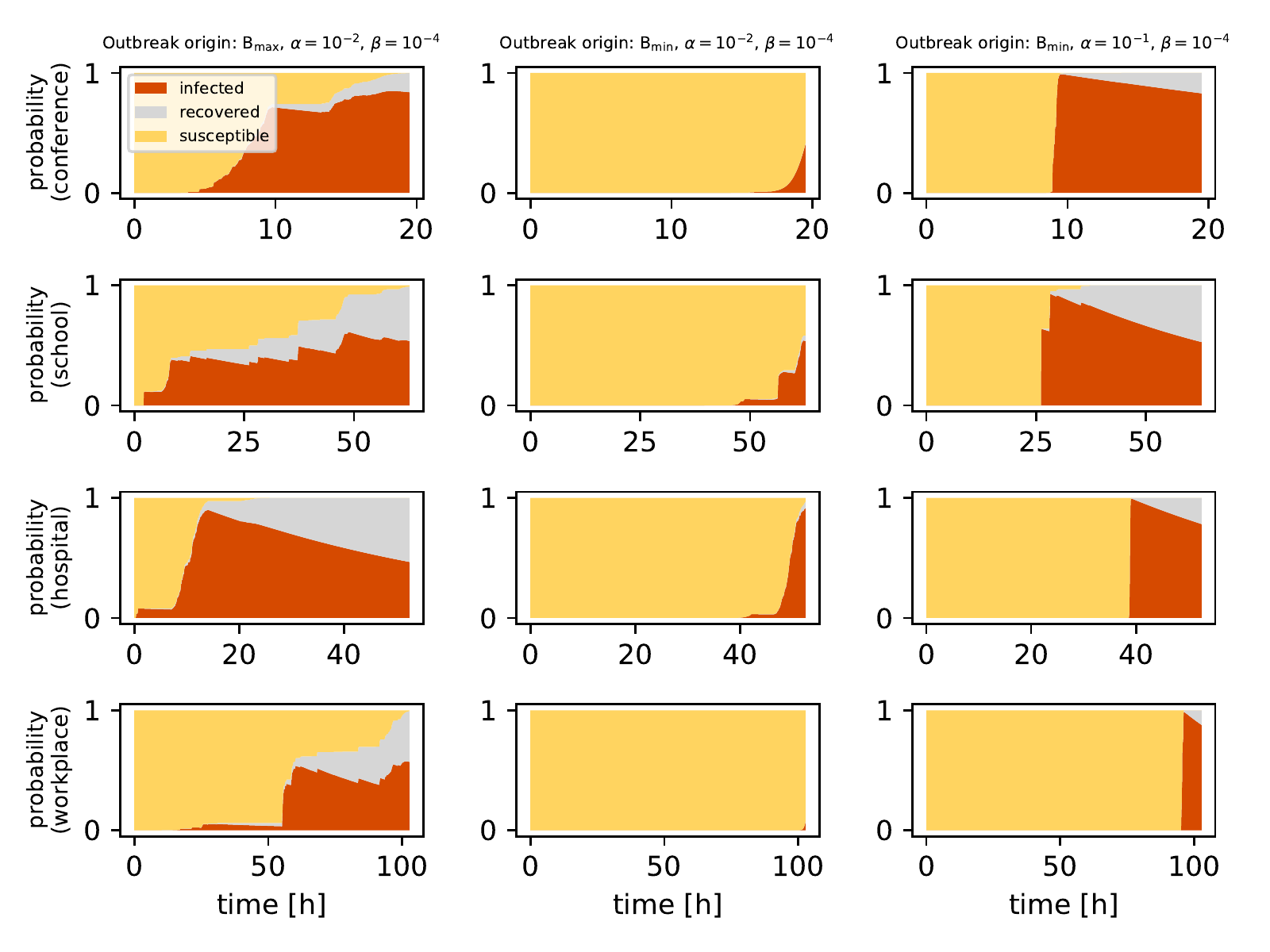}
\caption{Simulations of the SIR epidemic spreading when the origin of the outbreak of infection is the node that has: (a) the maximum broadcast $B_{\mbox{max}}$ (1st column); (b)  the minimum broadcast  $B_{\mbox{min}}$ (2nd column).  The probabilities of infection and recovery are set to $\alpha = 10^{-2}$ and $\beta = 10^{-4}$, respectively. The third column (c) shows the case in which the origin of the outbreak is the node with minimum broadcast but the probability of infection is ten times greater. In the figure are plotted the probability that an arbitrarily selected node is susceptible, infected or recovered.}
\label{fig_probsir}
\end{center}
\end{figure}

Fig.~\ref{fig_probsir} shows the simulations of an epidemic outbreak from a single initially infected node. The figure displays the probability that another arbitrarily selected node is in the susceptible, infected or recovered state. The first (a) and second (b) columns refer to the cases in which the origin of the outbreak of infection is the node with maximum and minimum broadcast, respectively. In these cases the probability of infection and recovery are set to $\alpha = 10^{-2}$ and $\beta = 10^{-4}$, respectively. The third column (c) refers to the case in which the origin of the outbreak is the node with minimum broadcast but the probability of infection is ten times greater than in the first two cases. It is clear that if the origin of the outbreak is the node with maximum broadcast the epidemic begins to spread much earlier than in the other cases, which support the fact that a node that has high broadcast quickly triggers the spread of the epidemic. On the other hand, the diffusive process is delayed or inhibited if the node seed of infection has low broadcast even if the probability of infection increases notably. In the case (b) for the Workplace network we observe that the probability that an arbitrarily selected node is infected is negligible.

\begin{figure}[!htbp]
\begin{center}
    \includegraphics[width=\textwidth]{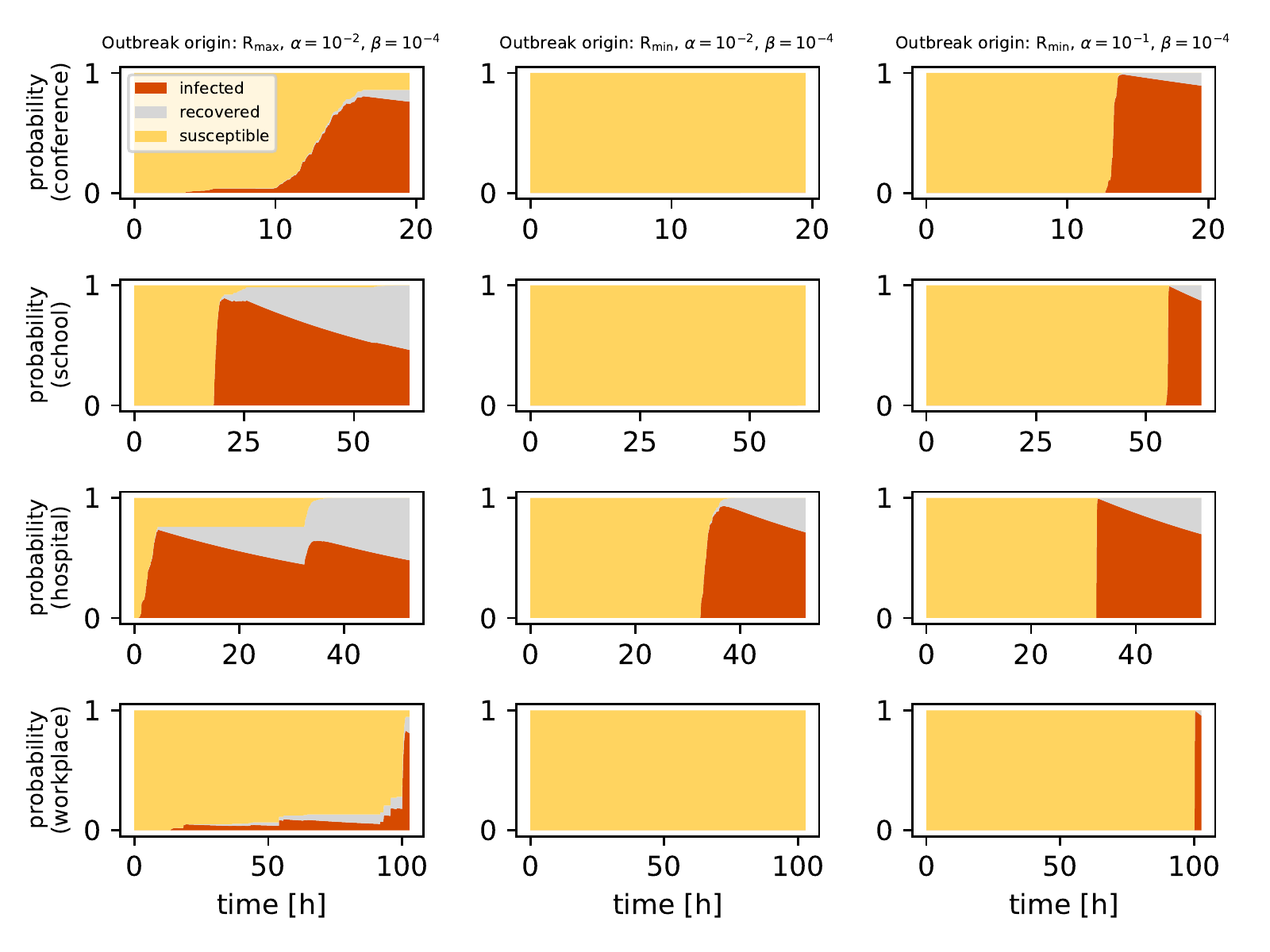}
\caption{The figure shows the results of the same SIR simulations of Fig.~\ref{fig_probsir} but performed over the time-reversed networks and with the role of broadcast replaced by that of receive $B_{\mbox{max}} \rightarrow R_{\mbox{max}}$, $B_{\mbox{min}} \rightarrow R_{\mbox{min}}$ for the node that is the origin of the outbreak.}
\label{fig_probsirtimerev}
\end{center}
\end{figure}

To study the propensity of a node to be a target of infection we look at its receiving capacity. To this end we run the same SIR simulations on the networks obtained by reversing the arrow of time and replacing the role of broadcast with that of receive $B_{\mbox{max}} \rightarrow R_{\mbox{max}}$, $B_{\mbox{min}} \rightarrow R_{\mbox{min}}$. Fig.~\ref{fig_probsirtimerev} shows the results of these simulations. This time the epidemic spreads early if the origin of the outbreak of infection is the node that has maximum receive, while it is delayed and inhibited if the seed of infection is the node that has minimum receive. We observe that in the case (b) there are even three networks, that is, Conference, School and Workplace for which the probability that an arbitrarily selected node is infected is zero.

\begin{figure}[!htbp]
\begin{center}
    \includegraphics[width=\textwidth]{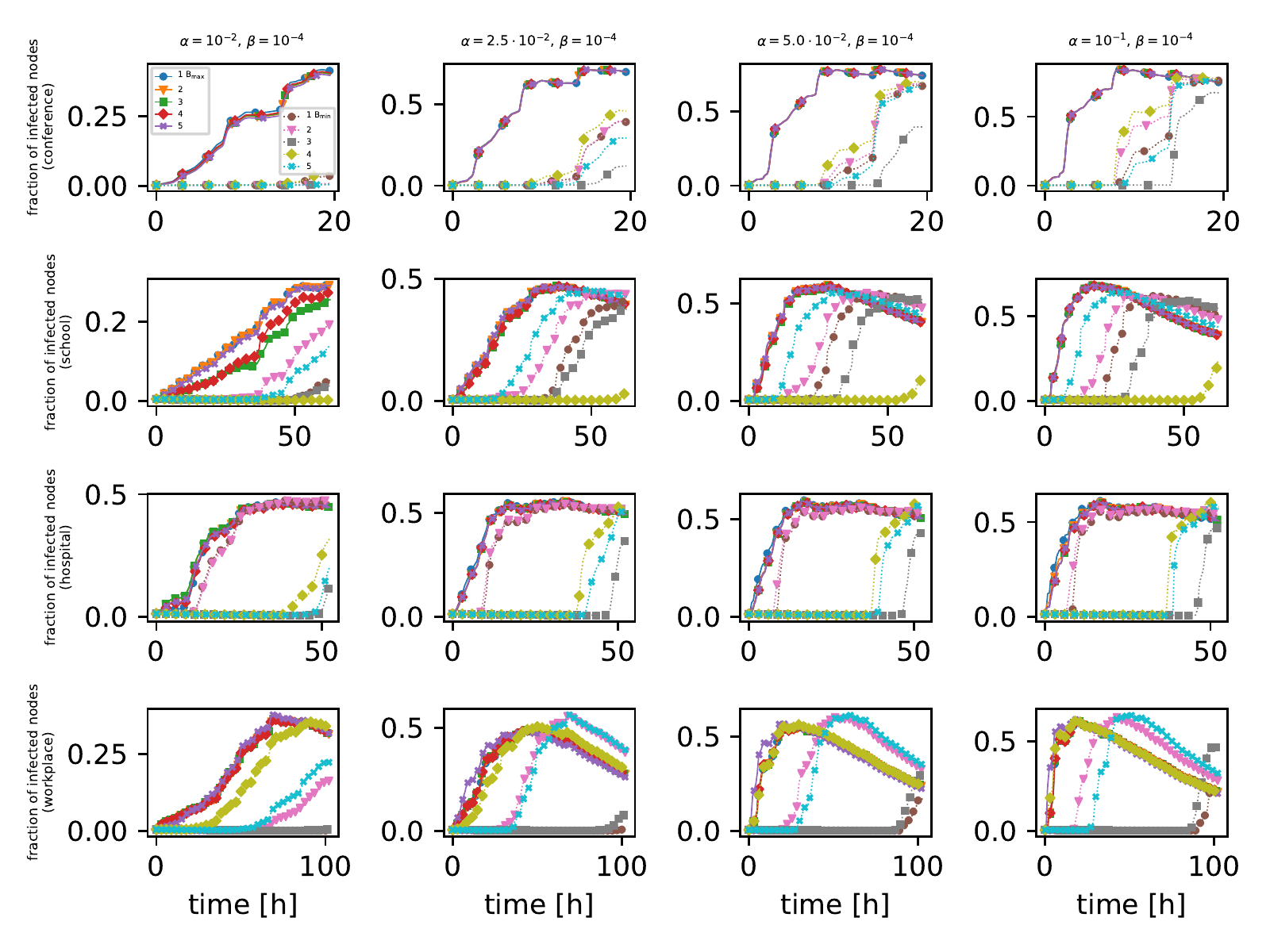}
\caption{The figure shows the fraction of infected nodes as a function of time for the SIR simulations when the infection probability is $\alpha = \{1.0, 2.5, 5.0, 10.0\} \times 10^{-2}$ and the recovery probability is $\beta = 10^{-4}$. Two groups of nodes were selected: \textbf{\RNum{1}}) five nodes ranked in descending order from $B_{\mbox{max}}$; \textbf{\RNum{2}}) five nodes ranked in ascending order from $B_{\mbox{min}}$ and each of these nodes was chosen as the single origin of the outbreak infection. Solid (dotted) lines are numbered according to the rank of nodes in group  \textbf{\RNum{1}} (\textbf{\RNum{2}}).}
\label{fig_meaninfecbcvara}
\end{center}
\end{figure}

To make sure that we are able to distinguish clearly between nodes with high and low communicability even when $\alpha$ and $\beta$ vary, we compare the evolution of the epidemic spreading in two situations. We consider two groups of nodes: \textbf{\RNum{1}})  five nodes ranked in descending order from $B_{\mbox{max}}$ and  \textbf{\RNum{2}}) five nodes ranked in ascending order from $B_{\mbox{min}}$. Each of these nodes was chosen as the single seed of infection, and the SIR simulations were performed using a wider range of $\alpha$ and $\beta$ values. Fig.~\ref{fig_meaninfecbcvara} shows the fraction of infected nodes as a function of time when the probability of infection $\alpha$ varies from $10^{-2}$ to $10^{-1}$ being $\beta = 10^{-4}$. Solid lines describe the evolution of the epidemic when the origin of the outbreak is a node of group \textbf{\RNum{1}} and dotted lines refer to the case in which the origin of the outbreak is a node of group \textbf{\RNum{2}}. For all examined networks we observe that the fraction of infected nodes grows much faster if the origin of infection is a node of group \textbf{\RNum{1}}. 

\begin{figure}[!htbp]
\begin{center}
    \includegraphics[width=\textwidth]{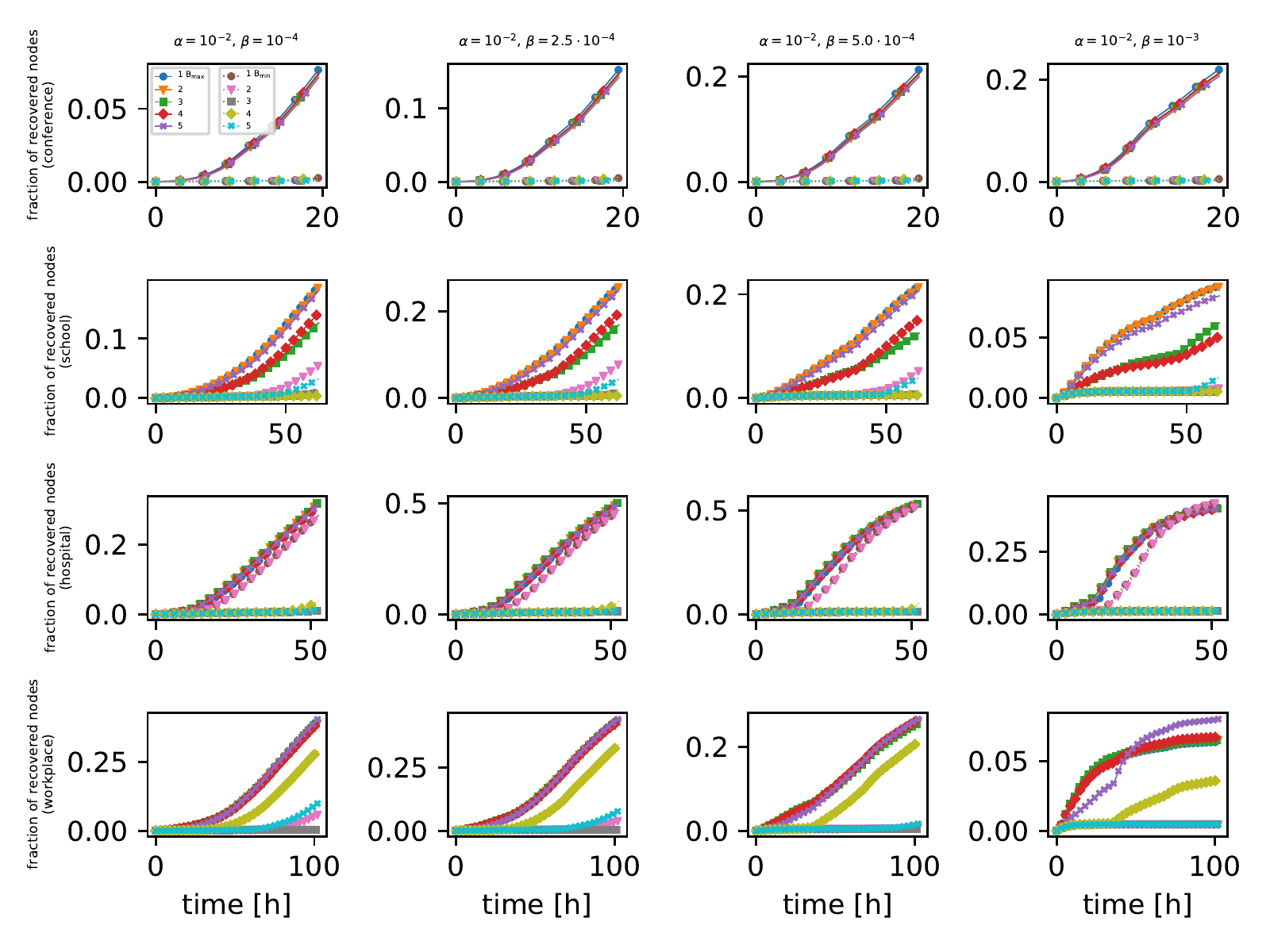}
\caption{The figure shows the fraction of recovered nodes as a function of time for the SIR simulations when the recover probability is $\beta = \{1.0, 2.5, 5.0, 10.0\} \times 10^{-4}$ and the infection probability is $\alpha = 10^{-2}$. Solid and dotted lines refer to the same two settings described in the caption of Fig.~\ref{fig_meaninfecbcvara} as regards the choice of the origin of the outbreak.}
\label{fig_meanrecbcvarb}
\end{center}
\end{figure}

Fig.~\ref{fig_meanrecbcvarb} shows the fraction of recovered nodes as a function of time when the recover probability $\beta$ varies from $10^{-4}$ to $10^{-3}$ being $\alpha = 10^{-2}$. Also in this case we note a clear distinction: the fraction of recovered nodes grows much faster when the origin of the outbreak infection is a node of group \textbf{\RNum{1}} rather than \textbf{\RNum{2}}.

\begin{figure}[!htbp]
\begin{center}
    \includegraphics[width=\textwidth]{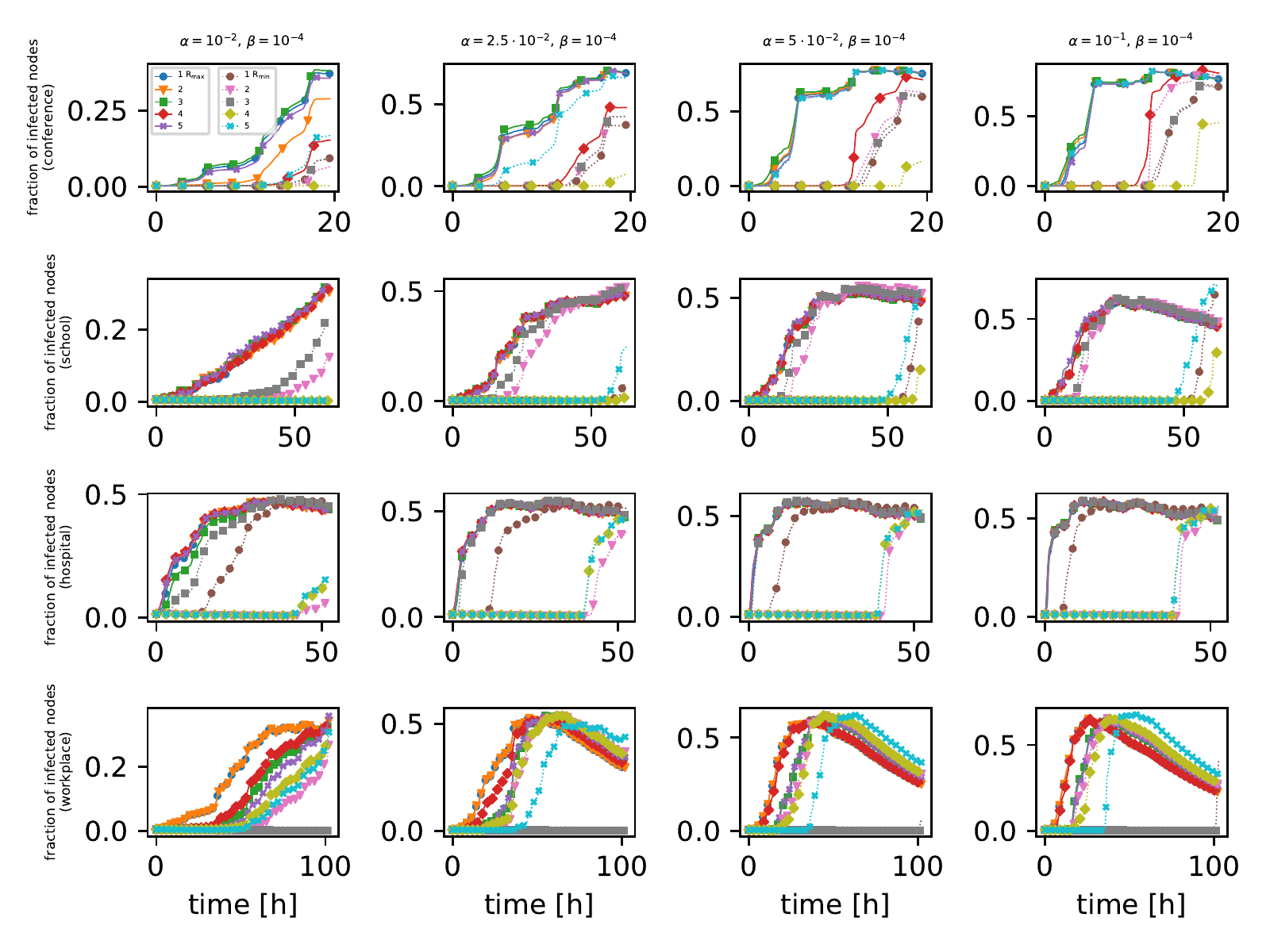}
\caption{The figure shows the results of the same SIR simulations described in Fig.~\ref{fig_meaninfecbcvara} but performed over the time-reversed networks and with the role of broadcast replaced by that of receive $B_{\mbox{max}} \rightarrow R_{\mbox{max}}$, $B_{\mbox{min}} \rightarrow R_{\mbox{min}}$ for the node that is the origin of the outbreak.}
\label{fig_meaninfecrcvara}
\end{center}
\end{figure}

\begin{figure}[!htbp]
\begin{center}
    \includegraphics[width=\textwidth]{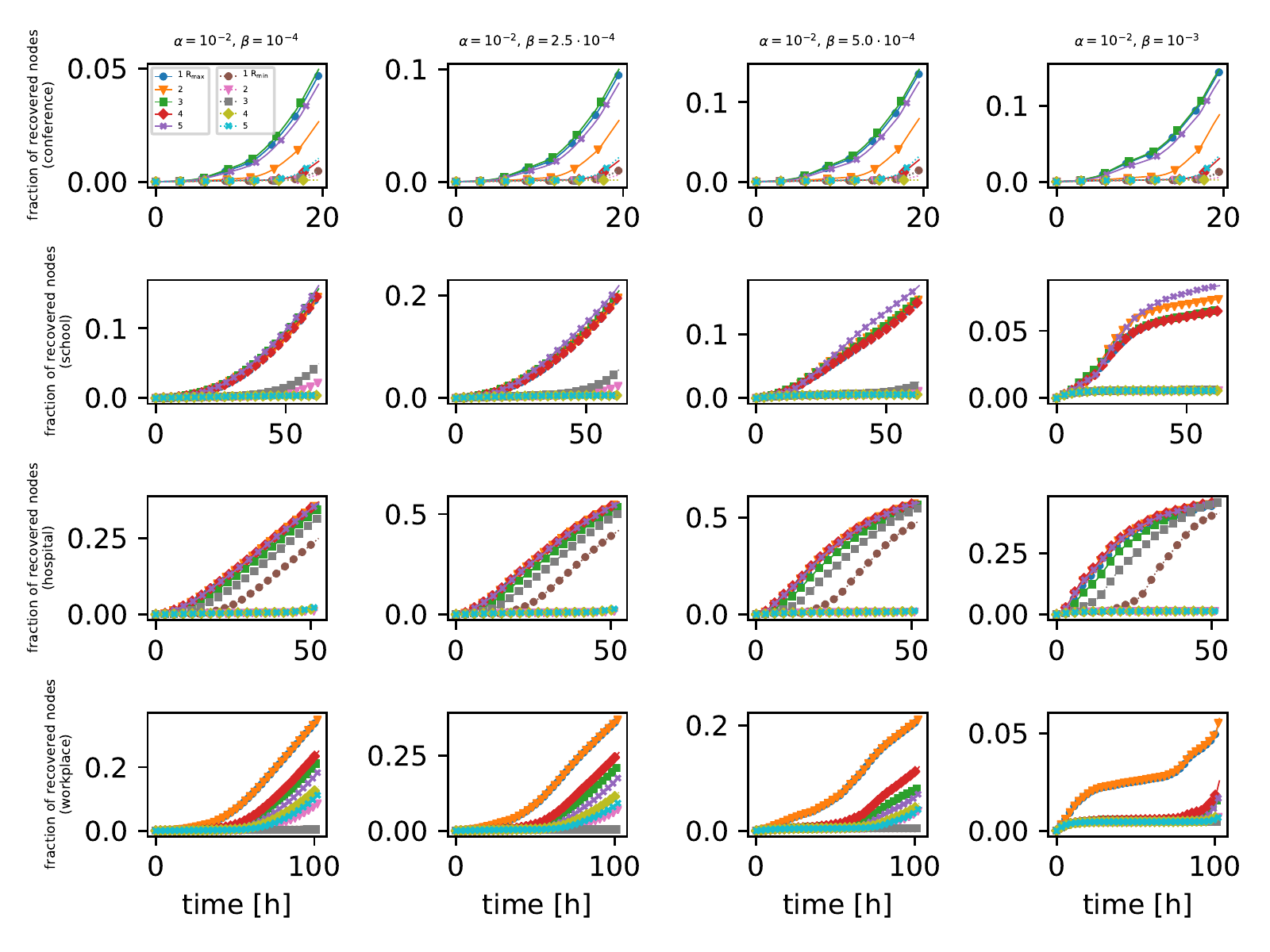}
\caption{The figure shows the results of the same SIR simulations described in Fig.~\ref{fig_meanrecbcvarb} but performed over the time-reversed networks and with the role of broadcast replaced by that of receive $B_{\mbox{max}} \rightarrow R_{\mbox{max}}$, $B_{\mbox{min}} \rightarrow R_{\mbox{min}}$ for the node that is the origin of the outbreak.}
\label{fig_meanrecrcvarb}
\end{center}
\end{figure}

To verify the correctness of the discrimination between nodes with high and low receive when $\alpha$ and $\beta$ vary, we use the technique of reversing the arrow of time and replacing the role of broadcast with that of receive.  Fig.~\ref{fig_meaninfecrcvara} and~\ref{fig_meanrecrcvarb} show the results of the same SIR simulations described in Fig.~\ref{fig_meaninfecbcvara} and~Fig.~\ref{fig_meanrecbcvarb}, respectively but performed on the networks obtained by reversing the chronological order of the snapshots and with the replacement $B_{\mbox{max}} \rightarrow R_{\mbox{max}}$, $B_{\mbox{min}} \rightarrow R_{\mbox{min}}$ for the node that is the origin of the outbreak. We observe that the fractions of infected and recovered nodes grows much faster when the origin of the outbreak is a node with high rather than low receive.  

These results show the effectiveness of the proposed method in clearly distinguishing high from low communicability nodes. Moreover, the greatest transmission and reception capacity of the nodes with high communicability, compared to those with low communicability, is proved by the fact that the epidemic spreads faster and on a larger scale, both in the original and in the time-reversed networks, when the source of infection is a node of the first type.  

\newpage

\section{Conclusion}\label{sec_conc}
We presented a method aimed to compute the communicability of nodes in temporal networks. An element $\mathbf{\mathcal{C}}_{ij}$ of the communicability matrix quantify how well information can be transmitted through causal (time-respecting) paths from a node $v_i$ to a node $v_j$ during the period of observation of the network. A path that has high communication efficiency starts as soon as possible after the initial instant of the period of observation, has the shortest duration and traverses the fewest intermediate nodes. The communicability matrix provides all possible chronological ordered products of adjacency matrices of the network snapshots and by means of a damping procedure favors paths with high communication efficiency and suppress those less  efficient. The damping procedure uses a weight function which depends on time and is monotonically increasing. 
We heuristically found that the logarithm function is a suitable choice but other functions may be adopted. The propensity of a node to transmit information  to other nodes is quantified by its broadcast and, similarly, its propensity to collect information is quantified by its receive and both can be computed as row and column sums of $\mathbf{\mathcal{C}}$, respectively. The chronological order of the product of adjacency matrices causes $\mathbf{\mathcal{C}}$ to be not symmetric even if so are the single adjacency matrices and reflects the fact that the arrow of time induces a direction in the communication flow. Transposing $\mathbf{\mathcal{C}}$ is equivalent to swapping the roles of broadcast and receive and, by construction, to reversing the chronological order of the products of adjacency matrices and, therefore, the arrow of time. 
We computed the communicability of four real-world networks of human proximity contacts and clearly identified the nodes with high communicability. We proved the accuracy of the method in distinguishing high from low communicability nodes by studying the spread of an epidemic in the networks  with the susceptible-infected-recovered model. The results of the simulations showed that the epidemic spreads early and on a larger scale, both in the original and in the time-reversed networks,  if the origin of infection is a node with high communicability while it is delayed and inhibited if the seed of infection is a node with low communicability.

The method presented in this work to compute the communicability matrix is based on simple linear algebra operations such as matrix sums and products and can be easily implemented on a computer. In this respect, the method is simpler than other proposed techniques based on resolvents, which require spectral analysis and may be computationally infeasible if the number of network snapshots and their size increase too much. 

\section{Acknowledgments}
The computing resources and the related technical support used for this work have been provided by CRESCO/ENEAGR\-ID High Performance Computing infrastructure and its staff~\cite{eneagrid}. CRESCO/ENEAGRID High Performance Computing infrastructure is funded by ENEA, the Italian National Agency for New Technologies, Energy and Sustainable Economic Development and by Italian and European research programmes, see https://www.eneagrid.enea.it for information.

\bibliography{communicabilitybib}

\end{document}